\documentclass[twocolumn,usenatbib]{aa}  
\usepackage{graphicx}
\usepackage{epsfig,graphicx,rotating,portland,fancyheadings,longtable}
\usepackage{multirow}
\usepackage{txfonts}

\def\ls{{_<\atop^{\sim}}}

\usepackage{amstext}

\begin{document}

\title{SN 2013dx associated with GRB\,130702A: a detailed photometric
  and spectroscopic monitoring and a study of the environment 
%
%
\thanks{Based on observations
    collected at the Italian 3.6-m Telescopio Nazionale Galileo (TNG),
    operated on the island of La Palma by the Fundacion Galileo
    Galilei of the INAF (Instituto Nazionale di Astrofisica) at the
    Spanish Observatorio del Roque de los Muchachos of the Instituto
    de Astrofisica de Canarias under program A27TAC 5, and at the
    European Southern Observatory, ESO, the VLT/Antu telescope,
    Paranal, Chile, proposal code: 291.D-5032(A).}}

\author{V. D'Elia$^{1,2}$, E. Pian$^{3,4}$, A. Melandri$^{5}$,
  P. D'Avanzo$^{5}$, M. Della Valle$^{6,7}$,
  P.A. Mazzali$^{8,9,10}$, S. Piranomonte$^{1}$,
  G. Tagliaferri$^{5}$, L.A. Antonelli$^{1,2}$, F. Bufano$^{11,12}$,
  S. Covino$^{5}$, D. Fugazza$^{5}$, D. Malesani$^{13}$, P. M\o ller$^{14}$,
  E. Palazzi$^{3}$ }

\institute
{$^{1}$INAF-Osservatorio Astronomico di Roma, Via Frascati 33, I-00040 Monteporzio Catone, Italy\\
$^{2}$ASI-Science Data Center, Via del Politecnico snc, I-00133 Rome, Italy\\
$^{3}$INAF - Istituto di Astrofisica Spaziale e Fisica Cosmica, Via P. Gobetti 101,  40129 Bologna, Italy \\
$^{4}$Scuola Normale Superiore, Piazza dei Cavalieri 7, 56126, Pisa, Italy\\
$^{5}$INAF-Osservatorio Astronomico di Brera, via E. Bianchi 46, 23807 Merate (LC), Italy\\  
$^{6}$INAF - Osservatorio Astronomico di Capodimonte, Salita Moiariello 16, I-80131 Napoli, Italy\\
$^{7}$ICRANET, Piazza della Republica 10, Pescara, I-65122, Italy\\
$^{8}$ARI, Liverpool John Moores University, IC2 Liverpool Science Park 146 Brownlow Hill, Liverpool, L3 5RF, UK\\
$^{9}$INAF - Osservatorio Astronomico di Padova, Vicolo dell’Osservatorio 5, I-35122 Padova, Italy\\
$^{10}$Max-Planck Institute for Astrophysics, Garching, Karl-Schwarzschild-Str. 1, Postfach 1317, D-85741 Garching, Germany\\
$^{11}$Millennium Institute of Astrophysics, Casilla 36-D, Santiago, Chile\\
$^{12}$Departamento de Ciencias Fisicas, Universidad Andres Bello, Avda. Republica 252, Santiago, Chile\\
$^{13}$Dark Cosmology Center, Niels Bohr Institute, University of Copenhagen, Juliane Maries Vej 30, 2100 Copenhagen, Denmark  \\
$^{14}$European Southern Observatory, Karl-Schwarzschild-Strasse 2, D-85748 Garching bei M\"unchen, Germany\\
}

  \abstract 
  {}
  {Long-duration gamma-ray bursts (GRBs) and broad-line, type Ic
    supernovae (SNe) are strongly connected. We aim at characterizing
    SN 2013dx, which is associated with GRB\,130702A, through sensitive and
    extensive ground-based observational campaigns in the optical-IR
    band.}
  {We monitored the field of the Swift GRB 130702A (redshift z =
    0.145) using the 8.2 m VLT, the 3.6 m TNG and the 0.6 m REM
    telescopes during the time interval between 4 and 40 days after
    the burst. Photometric and spectroscopic observations revealed the
    associated type Ic SN 2013dx. Our multiband photometry allowed
    constructing a bolometric light curve.}
  { The bolometric light curve of SN 2013dx resembles that of 2003dh
    (associated with GRB\,030329), but is $\sim$10\% faster and
    $\sim$25\% dimmer.  From this we infer a synthesized $^{56}$Ni
    mass of $\sim$0.2 M${_\odot}$.  The multi-epoch optical
    spectroscopy shows that the SN 2013dx behavior is best matched by
    SN 1998bw, among the other well-known low-redshift SNe associated with
    GRBs and XRFs, and by SN 2010ah, an energetic type Ic SN not
    associated with any GRB. The photospheric velocity of the ejected
    material declines from $\sim 2.7\times 10^{4}$ km s$^{-1}$ at $8$
    rest frame days from the explosion, to $\sim 3.5\times 10^{3}$ km
    s$^{-1}$ at $40$ days. These values are extremely close to those
    of SN1998bw and 2010ah. We deduce for SN 2013dx a kinetic energy
    of $\sim 35 \times 10^{51}$ erg and an ejected mass of $\sim 7$
    M${_\odot}$. This suggests that the progenitor of SN2013dx had a
    mass of $\sim$25-30 M$_\odot$, which is 15-20\% less massive than that
    of SN 1998bw.

    Finally, we studied the SN 2013dx environment through
    spectroscopy of the closeby galaxies: $9$ out of the $14$
    inspected galaxies lie within $0.03$ in redshift from $z=0.145$,
    indicating that the host of GRB\,130702A/SN 2013dx belongs to a
    group of galaxies, an unprecedented finding for a GRB-associated
    SN and, to our knowledge, for long GRBs in general.

}
{}

\authorrunning {D'Elia et al.}
\titlerunning {SN 2013dx}

\maketitle
%

\section{Introduction}

The connection between gamma-ray bursts and supernovae has been firmly
established on the basis of a handful of nearby events ($z < 0.3$) for
which a decent spectroscopic monitoring was possible (Galama et
al. 1998; Patat et al. 2001; Hjorth et al. 2003; Stanek et al. 2003;
Malesani et al. 2004; Pian et al. 2006; Bufano et al. 2012), which
enabled deriving the physical properties of the SNe (Mazzali
et al. 2006a,b; Woosley \& Bloom 2006; Hjorth \& Bloom 2012).

In these few cases, observations have revealed that SNe accompanying
GRBs are explosions of bare stellar cores, that is, their progenitors
(whose estimated mass is higher than $\sim$20 M$_\odot$) have lost all
their hydrogen and helium envelopes before collapse
(a.k.a. supernovae of type Ic).  A stripped star of Wolf-Rayet type
(Crowther 2007) appears indeed more suited to favor the propagation
of a relativistic jet.  However, many unknowns still surround the
GRB-SN connection: the properties of their progenitors (e.g., their
multiplicity and metallicity, Podsiadlowski et al.  2010; Levesque et
al. 2014), the nature of their remnants (e.g., Woosley \& Heger 2012;
Mazzali et al. 2014), and the formation and propagation of the jets
through the stellar envelope (Zhang et al. 2003; Uzdensky \& MacFadyen
2006; Fryer et al. 2009; Lyutikov 2011; Bromberg et al. 2014).

The detailed analysis of the physics of the supernovae can elucidate
these problems, and accurate optical spectroscopy is a fundamental tool
to this aim.  It is well known, in fact, that light curve models alone
do not have the ability to uniquely determine all supernova
parameters. In particular, while the mass of $^{56}$Ni can be
reasonably well established from the light-curve peak if the time of
explosion is known to within a good accuracy, or from the late,
exponential decline phase of the light curve if this is observed, the
two parameters that determine the shape of the light curve, the ejecta
mass $M_{ej}$ and the kinetic energy $E_K$, are degenerate (Eq.
2; Arnett 1982). Only the simultaneous use of light curve and spectra
can break this degeneracy (Mazzali et al. 2013, and references
therein). 

Time-resolved optical spectra of GRB-SNe can be acquired only
at $z \ls 0.3, $ however.  At higher redshift, the subtraction of the host galaxy
and afterglow components from the spectrum of the GRB optical
counterpart lowers the signal-to-noise ratio dramatically, so that the
resulting spectral residuals are noisy (e.g., Melandri et al. 2012;
2014). In fact, in the redshift range $0.3 \ls z \ls 1$ the current
generation of telescopes can test the SN-GRB connection only on the
basis of a single spectrum acquired at the epoch of maximum light
(Della Valle et al. 2003; Soderberg et al. 2005; Berger et al. 2011;
Sparre et al. 2011; Jin et al. 2013, Cano et al. 2014) or through the
detection of rebrightenings in the GRB afterglow light-curves, due to
emerging SNe (e.g. Bloom et al. 1999; Lazzati et al. 2001; Greiner et
al. 2003; Garnavich et al. 2003; Masetti et al. 2003; Zeh et al. 2004;
Gorosabel et al. 2005; Bersier et al. 2006; Della Valle et al. 2006;
Soderberg et al. 2006, 2007; Cobb et al. 2010; Tanvir et al. 2010;
Cano et al. 2011a,b; 2014). Finally, at redshift higher than one, even
the detection of a rebrightening is in general not possible.

SN~2013dx, associated with GRB\,130702A at $z=0.145$, represents a
remarkable new entry into the group of the most consistently
investigated SN-GRB events. Here, we present the results of our
extensive photometric and spectroscopic campaign, carried out with the
VLT, TNG, and REM telescopes, covering an interval of about $40$ days
after GRB detection.  In particular, we obtained 16 spectra spaced
apart by 2-3 days.

The paper is organized as follows: Section 2 summarizes the properties
of GRB\,130702A and its associated SN 2013dx as reported in the
literature; Sect. 3 introduces our dataset and illustrates the data
reduction process; Sect. 4 presents our results; in Sect. 5 we
draw our conclusions. We assume a cosmology with $H_0=70$ km s$^{-1}$
Mpc$^{-1}$, $\Omega_{\rm m} = 0.3$, $\Omega_\Lambda = 0.7$.

\section{GRB\,130702A/SN 2013 dx}

GRB\,130702A was detected by {\it Fermi}-LAT and -GBM instruments on
July 02, 2013 at 00:05 UT (Fermi trigger 394416326, Cheung et
al. 2013; Collazzi et al. 2013). In the following, this time
represents our $t0$.

The intermediate Palomar Transient Factory reported a possible
counterpart after following up the burst in the optical band at the
position RA(J2000) = $14h 29m 14.78s$ DEC(J2000) = +$15d 46'
26.4"$. The transient was $0.6"$ away from an SDSS faint source ($r =
23.01$) classified as a star, which was later identified as its host galaxy
(Singer et al. 2013; Kelly et al. 2013). A {\it Swift} target-of-opportunity observation was then activated, and analysis of the XRT
and UVOT data revealed a coincident, new X-ray source in the sky
(D'Avanzo et al. 2013a). The fading behavior was confirmed both in the
X-ray by subsequent XRT observations (D'Avanzo et al. 2013c) and in
the optical bands by several ground-based telescopes (see, e.g.,
Guidorzi et al. 2013; Xu et al. 2013; D'Avanzo et al. 2013b).

GRB\,130702A was also observed by Konus-Wind in the $20-1200$ keV band
(Golenetskii et al. 2013) and by INTEGRAL (Hurley et al. 2013) in the mm (Perley and Kasliwal 2013) and in the radio band (van der Horst
2013).

The redshift of the transient was reported to be $z=0.145$ (Mulchaey
et al. 2013a; D'Avanzo et al. 2013d; Mulchaey et al. 2013b). The
emerging of a new supernova associated with GRB\,130702A was detected
by several spectroscopic measurements (Schulze et al. 2013; Cenko et
al. 2013; D'Elia et al. 2013), and was named SN~2013dx (Schulze et al. 2013b,
D'Elia et al. 2013b).

\section{Observations and data reduction}

We observed the field of SN 2013dx in imaging and spectroscopic
modes using VLT, TNG, and REM. Details on the data acquisition and
reduction process are given below. Figure 1 shows the SN 2013dx
  field, with two nearby galaxies clearly visible. Details on the
  analysis of this field can be found in Sect. 4.6.


\begin{figure}
\centering
\includegraphics[angle=0,width=9cm]{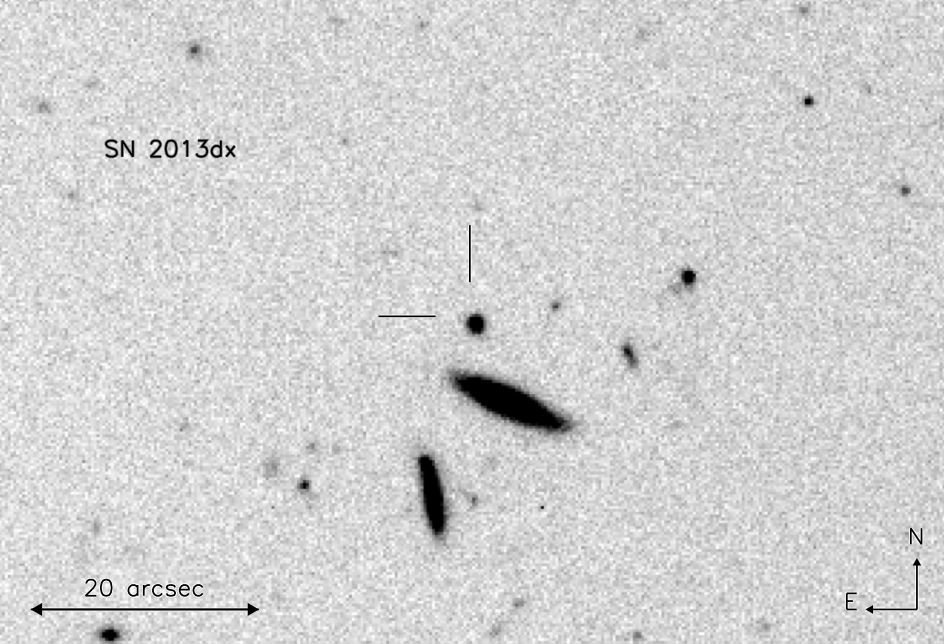}
\caption{SN 2013dx field acquired with VLT/FORS2 on July  27
  in the R band. The SN and host galaxy position is indicated by two
  perpendicular markers. Two companion galaxies are clearly visible
  southward. }
\end{figure}

\subsection{Photometry}

\subsubsection{TNG photometry}

TNG photometry was conducted using the DOLORES camera in imaging mode,
with the SDSS g,r,i,z and the Johnson U filters. Image reduction was
carried out by following the standard procedures: subtraction of an
averaged bias frame, division by a normalized flat frame.  Astrometry
was performed using the
USNOB1.0\footnote{http://www.nofs.navy.mil/data/fchpix/} catalogs.
Aperture photometry was made with the
Starlink\footnote{http://starlink.jach.hawaii.edu/starlink} PHOTOM
package. The aperture was set to $10$ pixels, which is equivalent to
$2.5"$.  The calibration was made against the SDSS catalog and
Landolt standard field stars (for the U filter). To minimize
any systematic effect, we performed differential photometry with
respect to local isolated and non-saturated standard stars selected
from these two catalogs.

The log of the TNG photometric observations can be found in Table
1.

\begin{table}[ht]
\caption{\bf Log of the TNG photometric observations}
{\footnotesize
\smallskip
\begin{tabular}{|l|c|c|c|c|c|}
\hline 

Date$^a$      &    t-t0 (d)$^b$     &   texp(s)$^c$    &   filt.$^d$          &    mag$^e$              &flux ($\mu$Jy)$^f$          \\
\hline
\hline
0705  &    $  3.89692$  &   $  2\times300$  &           U      &    $19.95\pm0.04$ &   $23.4\pm0.9$ \\
0720  &    $ 18.90617$  &   $  5\times120$  &           U      &    $21.27\pm0.26$ &   $7.0\pm1.9$ \\
0723  &    $ 21.90367$  &   $ 10\times120$  &           U      &    $21.94\pm0.27$ &   $3.7\pm1.0$ \\
0804  &    $ 33.89719$  &   $  5\times240$  &           U      &    $22.66\pm0.28$ &   $1.9\pm0.6$ \\
\hline                                                                                        
0703  &    $  1.96469$  &   $  2\times300$  &           g       &    $19.29\pm0.06$ &   $80.4\pm4.3$ \\
0705  &    $  3.88074$  &   $  1\times300$  &           g       &    $20.12\pm0.08$ &   $37.4\pm2.8$ \\
0714  &    $ 12.93708$  &   $  1\times300$  &           g       &    $20.33\pm0.12$ &   $30.9\pm3.5$ \\
0716  &    $ 14.92913$  &   $  1\times300$  &           g       &    $20.40\pm0.10$ &   $28.9\pm2.9$ \\
0717  &    $ 15.90906$  &   $  1\times300$  &           g       &    $20.38\pm0.04$ &   $29.6\pm1.0$ \\
0720  &    $ 18.89108$  &   $  1\times300$  &           g       &    $20.76\pm0.07$ &   $20.7\pm1.4$ \\
0723  &    $ 21.88002$  &   $  1\times300$  &           g       &    $21.05\pm0.10$ &   $16.0\pm1.6$ \\
\hline                                                                                               
0703  &    $  1.95541$  &   $  2\times300$  &           r       &    $19.10\pm0.01$ &   $91.9\pm1.2$ \\
0705  &    $  3.89015$  &   $  1\times300$  &           r       &    $19.84\pm0.02$ &   $46.6\pm0.8$ \\
0709  &    $  7.91513$  &   $  1\times300$  &           r       &    $19.91\pm0.05$ &   $43.5\pm2.1$ \\
0714  &    $ 12.94571$  &   $  1\times300$  &           r       &    $19.77\pm0.03$ &   $49.8\pm1.2$ \\
0716  &    $ 14.92508$  &   $  1\times300$  &           r       &    $19.76\pm0.03$ &   $50.2\pm1.2$ \\
0717  &    $ 15.92203$  &   $  1\times300$  &           r       &    $19.71\pm0.06$ &   $52.6\pm2.7$ \\
0720  &    $ 18.89967$  &   $  1\times300$  &           r       &    $19.85\pm0.04$ &   $46.1\pm1.6$ \\
0723  &    $ 21.89224$  &   $  1\times300$  &           r       &    $20.06\pm0.04$ &   $38.1\pm1.4$ \\
0804  &    $ 33.88716$  &   $  1\times300$  &           r       &    $20.92\pm0.06$ &   $17.3\pm1.0$ \\
\hline                                                                                      
0703  &    $  1.97366$  &   $  2\times300$  &           i       &    $18.97\pm0.05$ &   $101\pm4.5$ \\
0705  &    $  3.88548$  &   $  1\times300$  &           i       &    $19.89\pm0.04$ &   $43.5\pm1.7$ \\
0714  &    $ 12.94145$  &   $  1\times300$  &           i       &    $19.95\pm0.05$ &   $41.2\pm2.0$ \\
0716  &    $ 14.92082$  &   $  1\times300$  &           i       &    $19.95\pm0.06$ &   $41.2\pm2.4$ \\
0717  &    $ 15.92766$  &   $  1\times300$  &           i       &    $19.89\pm0.07$ &   $43.3\pm2.7$ \\
0720  &    $ 18.89542$  &   $  1\times300$  &           i       &    $19.82\pm0.06$ &   $46.5\pm2.8$ \\
0723  &    $ 21.88747$  &   $  1\times300$  &           i       &    $20.15\pm0.06$ &   $34.1\pm1.9$ \\
\hline                                                                                      
0705  &    $  3.90667$  &   $  5\times120$  &           z       &    $19.71\pm0.03$ &   $50.1\pm1.3$ \\
\hline
\hline

\end{tabular}
}
$^a$: Observation date (month-day)
$^b$: Time after the GRB (observer frame) 
$^c$: Exposure time
$^d$: Adopted filter 
$^e$: Magnitudes in AB system, except U-band magnitudes, which are Vega (not corrected for the Galactic extinction) 
$^f$: Fluxes (corrected for the Galactic extinction).
\end{table}

\subsubsection{VLT photometry}

VLT photometry was carried out using the FORS2 camera in imaging mode
with the Johnson U,B,R,I filters. Image reduction was performed using
the same standard techniques as were adopted for the TNG data. Images were
calibrated against Landolt standard field stars. Again, differential
photometry was performed to minimize any systematic effects. The
aperture was set to $10$ pixels, which is equivalent to $1.3"$.

The log of the VLT photometric observations can be found in Table
2. 

\begin{table}[ht]
\caption{\bf Log of the VLT photometric observations}
{\footnotesize
\smallskip
\begin{tabular}{|l|c|c|c|c|c|}
\hline
Date$^a$      &    t-t0 (d)$^b$     &   texp(s)$^c$    &   filt.$^d$          &    mag$^e$              &flux ($\mu$Jy)$^f$          \\
\hline 
\hline 
0710  &      8.032       &      $2\times60$      &      u   &    $  21.36\pm0.04$  &  $12.7\pm0.5$ \\
0711  &      9.974       &      $2\times60$      &      u   &    $  21.42\pm0.04$  &  $12.0\pm0.5$ \\
0713  &     11.966       &      $2\times60$      &      u   &    $  21.51\pm0.07$  &  $11.1\pm0.7$ \\
0716  &     14.974       &      $2\times60$      &      u   &    $  21.70\pm0.14$  &  $9.3 \pm1.3$ \\
0720  &     18.971       &      $2\times60$      &      u   &    $  22.53\pm0.60$  &  $4.3 \pm3.2$ \\
0722  &     20.972       &      $2\times60$      &      u   &    $  22.54\pm0.51$  &  $4.3 \pm2.6$ \\
0727  &     25.975       &      $6\times60$      &      u   &    $  23.06\pm0.12$  &  $2.7 \pm0.3$ \\
0730  &     28.980       &      $8\times60$      &      u   &    $  23.21\pm0.11$  &  $2.3 \pm0.2$ \\
0804  &     33.002       &      $8\times60$      &      u   &    $  23.20\pm0.16$  &  $2.3 \pm0.4$ \\
0806  &     35.966       &      $8\times60$      &      u   &    $  23.24\pm0.15$  &  $2.2 \pm0.3$ \\
0810  &     39.984       &      $8\times60$      &      u   &    $  23.46\pm0.21$  &  $1.8 \pm0.4$ \\
\hline                                                                                    
0710  &      8.034       &      $1\times60$      &      g   &    $  20.20\pm0.01$  &  $37.2\pm0.3$ \\
0711  &      9.976       &      $1\times60$      &      g   &    $  20.12\pm0.01$  &  $39.8\pm0.4$ \\
0713  &     11.968       &      $1\times60$      &      g   &    $  20.30\pm0.01$  &  $33.9\pm0.4$ \\
0716  &     14.976       &      $1\times60$      &      g   &    $  20.40\pm0.03$  &  $30.8\pm0.8$ \\
0720  &     18.973       &      $1\times60$      &      g   &    $  20.85\pm0.07$  &  $20.3\pm1.4$ \\
0722  &     20.974       &      $1\times60$      &      g   &    $  20.87\pm0.09$  &  $20.0\pm1.8$ \\
0727  &     25.979       &      $2\times60$      &      g   &    $  21.21\pm0.03$  &  $14.6\pm0.4$ \\
0730  &     28.987       &      $2\times60$      &      g   &    $  21.67\pm0.03$  &  $9.5 \pm0.2$ \\
0804  &     33.009       &      $2\times60$      &      g   &    $  21.85\pm0.04$  &  $8.1 \pm0.3$ \\
0806  &     35.973       &      $2\times60$      &      g   &    $  22.12\pm0.04$  &  $6.3 \pm0.2$ \\
0810  &     39.991       &      $2\times60$      &      g   &    $  22.17\pm0.06$  &  $6.0 \pm0.3$ \\
\hline                                                                                    
0710  &      8.035       &      $1\times60$      &      r   &    $  19.99\pm0.01$  &  $44.9\pm0.3$ \\
0710  &      8.040       &      $1\times60$      &      r   &    $  19.97\pm0.01$  &  $45.7\pm0.3$ \\
0711  &      9.977       &      $1\times60$      &      r   &    $  19.91\pm0.01$  &  $48.4\pm0.3$ \\
0711  &      9.982       &      $1\times60$      &      r   &    $  19.90\pm0.01$  &  $48.9\pm0.3$ \\
0713  &     11.969       &      $1\times60$      &      r   &    $  19.84\pm0.01$  &  $51.6\pm0.3$ \\
0716  &     14.977       &      $1\times60$      &      r   &    $  19.82\pm0.01$  &  $52.7\pm0.5$ \\
0720  &     18.974       &      $1\times60$      &      r   &    $  19.93\pm0.01$  &  $47.4\pm0.6$ \\
0722  &     20.968       &      $1\times60$      &      r   &    $  19.94\pm0.02$  &  $47.1\pm0.9$ \\
0727  &     25.981       &      $1\times60$      &      r   &    $  20.25\pm0.01$  &  $35.4\pm0.3$ \\
0730  &     28.989       &      $1\times60$      &      r   &    $  20.35\pm0.01$  &  $32.2\pm0.3$ \\
0804  &     33.011       &      $1\times60$      &      r   &    $  20.68\pm0.02$  &  $23.7\pm0.3$ \\
0806  &     35.975       &      $1\times60$      &      r   &    $  20.95\pm0.02$  &  $18.6\pm0.3$ \\
0810  &     39.994       &      $1\times60$      &      r   &    $  21.20\pm0.03$  &  $14.8\pm0.4$ \\
\hline                                                                                    
0710  &      8.037       &      $1\times60$      &      i   &    $  20.22\pm0.02$  &  $36.3\pm0.5$ \\
0711  &      9.979       &      $1\times60$      &      i   &    $  20.14\pm0.01$  &  $39.3\pm0.4$ \\
0713  &     11.971       &      $1\times60$      &      i   &    $  20.07\pm0.01$  &  $41.8\pm0.5$ \\
0716  &     14.978       &      $1\times60$      &      i   &    $  19.93\pm0.02$  &  $47.3\pm0.7$ \\
0720  &     18.976       &      $1\times60$      &      i   &    $  19.76\pm0.08$  &  $55.3\pm4.0$ \\
0722  &     20.977       &      $1\times60$      &      i   &    $  19.99\pm0.02$  &  $44.8\pm1.0$ \\
0727  &     25.982       &      $1\times60$      &      i   &    $  20.23\pm0.02$  &  $35.9\pm0.6$ \\
0730  &     28.990       &      $1\times60$      &      i   &    $  20.28\pm0.02$  &  $34.3\pm0.6$ \\
0804  &     33.013       &      $1\times60$      &      i   &    $  20.49\pm0.05$  &  $28.4\pm1.3$ \\
0806  &     35.977       &      $1\times60$      &      i   &    $  20.60\pm0.03$  &  $25.5\pm0.6$ \\
0810  &     39.995       &      $1\times60$      &      i   &    $  21.02\pm0.06$  &  $17.4\pm0.9$ \\
\hline
\hline

\end{tabular}
}
$^a$: Observation date (month-day)
$^b$: Time after the GRB (observer frame) 
$^c$: Exposure time
$^d$: Adopted filter 
$^e$: Magnitudes in AB system (not corrected for the Galactic extinction) 
$^f$: Fluxes (corrected for the Galactic extinction).
\end{table}

\subsubsection{REM photometry}

Optical and NIR observations were performed with the REM telescope
(Zerbi et al. 2001; Chincarini et al. 2003; Covino et al. 2004)
equipped with the ROSS2 optical imager and the REMIR NIR camera. The
ROSS2 instrument is able to observe simultaneously in the g, r, i, and
z SDSS filters.  Observations of GRB\,130702A/SN\,2013dx were carried
out over fifteen epochs between 2013 July 10 and Aug 13.  Image
reduction was carried out by following the same standard procedures as
for the TNG and VLT photometry, including differential photometry. The
aperture was set to $10$ pixels, which is equivalent to
$5"$. Astrometry was performed using the USNOB1.0 and the
2MASS\footnote{http://www.ipac.caltech.edu/2mass/} catalogs for the
optical and NIR frames, respectively.  The calibration was made
against the SDSS catalog for the optical filters and the 2MASS
catalog for NIR filters.

The log of the REM photometric observations can be found in Table 3,
where we only report the g and r photometry.  In the i band the S/N
was too low, and the SN 2013dx too much contaminated by the host
galaxy to obtain firm detections. In the z and H bands we did not
detect the SN down to typical upper limits of $z\sim 19.7$ (AB,
$3\sigma$) and $H\sim 18$ (Vega, $3\sigma)$. The columns are organized
in the same way as in Table 1.

\begin{table}[ht]
\caption{\bf Log of the REM photometric observations}
{\footnotesize
\smallskip
\begin{tabular}{|l|c|c|c|c|c|}
\hline 

Date$^a$      &    t-t0 (d)$^b$     &   texp(s)$^c$    &   filt.$^d$          &    mag$^e$              &flux ($\mu$Jy)$^f$          \\
\hline
\hline

0711  &     $9.04484 $   &   $9\times300$       &       g    &    $20.54\pm0.19$  &  $25.4\pm9.2$\\
0713  &     $11.05959$   &   $9\times300$       &       g    &    $20.56\pm0.08$  &  $25.1\pm3.6$\\
0715  &     $13.10455$   &   $9\times300$       &       g    &    $20.40\pm0.11$  &  $28.9\pm5.9$\\
0716  &     $14.10692$   &   $9\times300$       &       g    &    $20.45\pm0.13$  &  $27.6\pm6.9$\\
0718  &     $16.06110$   &   $9\times300$       &       g    &    $20.32\pm0.22$  &  $31.3\pm13$\\
0719  &     $17.09980$   &   $9\times300$       &       g    &    $21.56\pm0.41$  &  $10.0\pm8.7$\\
0725  &     $22.99765$   &   $9\times300$       &       g    &    $21.36\pm0.13$  &  $12.0\pm3.0$\\   
\hline                                                                                                    
0711  &     $ 9.04484$   &   $9\times300$       &       r    &    $19.96\pm0.09$  &  $43.5\pm4.9$\\
0713  &     $11.05959$   &   $9\times300$       &       r    &    $19.91\pm0.07$  &  $45.3\pm4.1$\\
0715  &     $13.10455$   &   $9\times300$       &       r    &    $19.75\pm0.10$  &  $52.6\pm6.7$\\
0716  &     $14.10692$   &   $9\times300$       &       r    &    $19.62\pm0.14$  &  $59.5\pm10$\\
0718  &     $16.06110$   &   $9\times300$       &       r    &    $19.81\pm0.16$  &  $49.8\pm10$\\
0719  &     $17.09980$   &   $9\times300$       &       r    &    $19.88\pm0.16$  &  $46.6\pm9.5$\\
0725  &     $22.99765$   &   $9\times300$       &       r    &    $20.01\pm0.15$  &  $41.6\pm7.4$\\
\hline
\hline
\end{tabular}
}
$^a$: Observation date (month-day)
$^b$: Time from the GRB (observed frame) 
$^c$: Exposure time
$^d$: Adopted filter 
$^e$: Magnitudes in AB system (not corrected for the Galactic extinction) 
$^f$: Fluxes (corrected for the Galactic extinction).
\end{table}


\begin{figure}
\centering
\includegraphics[angle=0,width=9cm]{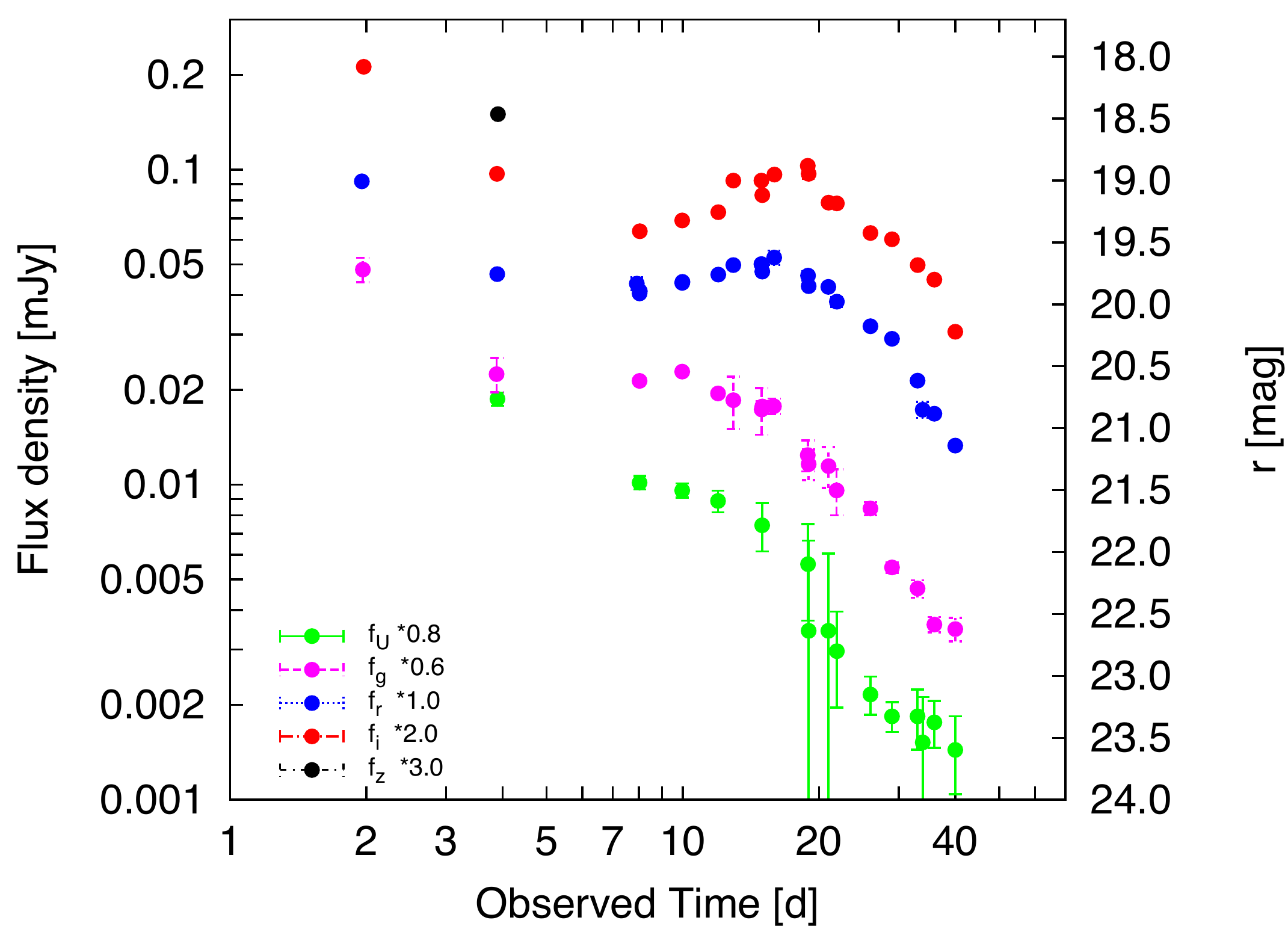}
\caption{GRB\,130702A/SN2013dx optical and near-infrared light
  curves. 
  The time origin $t=0$ coincides with the GRB explosion time. }
\end{figure}

\subsection{Spectroscopy}

\subsubsection{TNG spectroscopy}

TNG spectroscopy was carried out using the DOLORES camera in slit
mode, with the LR-B grism. This configuration covers the spectral
range $3000-8430$\AA\, with a resolution of $\lambda/\Delta\lambda = 585$
for a slit width of $1$'' at the central wavelength $5850$\AA. 
Spectra were acquired at five epochs. A slit width of $1$'' was used in
all but the first observation, for which a $1.5$'' slit was adopted,
owing to a seeing angle higher than $1$''. The slit position angle was
set to the parallactic angle in all our observations. Table 4
contains a summary of the TNG spectroscopic observations.

The spectra were extracted using standard procedures (bias and
background subtraction, flat fielding, wavelength and flux
calibration) under the packages
ESO-MIDAS\footnote{http://www.eso.org/projects/esomidas/} and
IRAF\footnote{http://iraf.noao.edu/}. Ne-Hg or helium lamps and
spectrophotometric stellar spectra acquired in the same nights as the
target were used for wavelength and flux calibration.

To account for slit losses, we matched the flux-calibrated spectra with
our multiband TNG photometry acquired  in the same nights.

\begin{table}[ht]
\caption{\bf Log of the TNG spectroscopic observations}
\centering
{\footnotesize
\smallskip
\begin{tabular}{|l|c|c|c|c|}
\hline

Epoch & obs. date  & exp. time (s)     & S/N   & slit width (arcsec)    \\
\hline
\hline
1TS      & 09 Jul   & $2000$             & $6-10$ & $1.5$\\
\hline
2TS      & 14 Jul   & $1800$             & $6-12$ & $1.0$\\
\hline
3TS      & 16 Jul   & $1800$             & $4-6$ &  $1.0$\\
\hline
4TS      & 17 Jul   & $1800$             & $4-6$ &  $1.0$\\
\hline
5TS      & 20 Jul   & $2\times1800$      & $6-8$ &  $1.0$\\
\hline
\hline

\end{tabular}
}
\end{table}

\subsubsection{VLT spectroscopy}

VLT spectroscopy was carried out using the FORS2 camera in slit mode,
with the 300V grism. This configuration covers the spectral range
$3300-9500$ \AA\, with a resolution of $\lambda/\Delta\lambda = 440$
for a slit width of $1$'' at the central wavelength $5900$ \AA. Eleven
spectroscopic epochs were acquired. A slit width of $1$'' was used in
all observations. The slit position angle was set to different values
in each observation to place different field galaxies in the
slit to study the SN 2013dx surroundings (see Sect. 4.2 for
details). Table 5 gives a summary of the VLT spectroscopic
observations.

As for the TNG data, the spectra were extracted using the packages
ESO-MIDAS and IRAF. A He+Ag/Cd+Ar lamp and spectrophotometric stars
acquired the same night of the target were used for wavelength and
flux calibration. For a few epochs, some
spectrophotometric stars were not observed in the same night as the
target. In these cases, the flux calibration was performed using
archival data. In none of these epochs, however, was the difference beween
the observation time of the scientific files and the adopted standard
longer than three days.

To account for slit losses, we checked the flux-calibrated spectra
against our simultaneous VLT multiband photometry.

\begin{table}[ht]
\caption{\bf Log of the VLT spectroscopic observations}
\centering
{\footnotesize
\smallskip
\begin{tabular}{|l|c|c|c|c|}
\hline

Epoch & obs. date  & exp. time (s)     & S/N      & PA (CCW from N)    \\
\hline
\hline
1VS      & 09 Jul   & $2\times900$      & $15-35$& $22.0\deg$ \\
\hline
2VS      & 11 Jul   & $900$             & $25-40$& $-14.2\deg$\\
\hline
3VS      & 13 Jul   & $900$             & $15-30$& $129.8\deg$\\
\hline
4VS      & 16 Jul   & $900$             & $6-13$ & $80.5\deg$\\
\hline
5VS      & 20 Jul   & $900$             & $3-5$  & $93.4\deg$ \\
\hline
6VS      & 22 Jul   & $900$             & $2-4$  & $59.0\deg$\\
\hline
7VS      & 27 Jul   & $900$             & $5-15$& $82.5\deg$\\
\hline
8VS      & 30 Jul   & $900$             & $6-10$& $22.0\deg$\\
\hline
9VS      & 03 Aug   & $900$             & $5-12$& $82.5\deg$\\
\hline
10VS     & 06 Aug   & $2\times1800$     & $9-22$& $50.9\deg$\\
\hline
11VS     & 10 Aug   & $2\times900$      & $3-6$ & $08.8\deg$ \\

\hline
\hline

\end{tabular}
}
\end{table}

\section{Results}

\subsection{Optical light curves}

Our optical photometric data, only corrected for Galactic extinction,
are shown in Fig. 2.  The SN component starts to emerge and dominate
its host galaxy and GRB optical afterglow at about $\text{five}$ days after
the GRB explosion. We fit empirical curves to the data past
day five to determine the peak times of SN2013dx in the different
bands. The results in the rest frame are $T_{peak,i} = 16 \pm 2$
days, $T_{peak,r} = 15 \pm 1$ days, $T_{peak,g} = 10 \pm 2$ days, and
$T_{peak,U} = 8 \pm 1$ days.

SN 2013dx is one of the few cases in which a SN is observed in the U
band.  The U peak is observed very early, at about one week in the
rest-frame. As commonly observed in other SNe, the peak occurs at
later times while moving to redder bands. As an example, the I-band
peak is observed about a week later than that in the U-band.


\begin{figure*}
\centering
\includegraphics[angle=0,width=18cm]{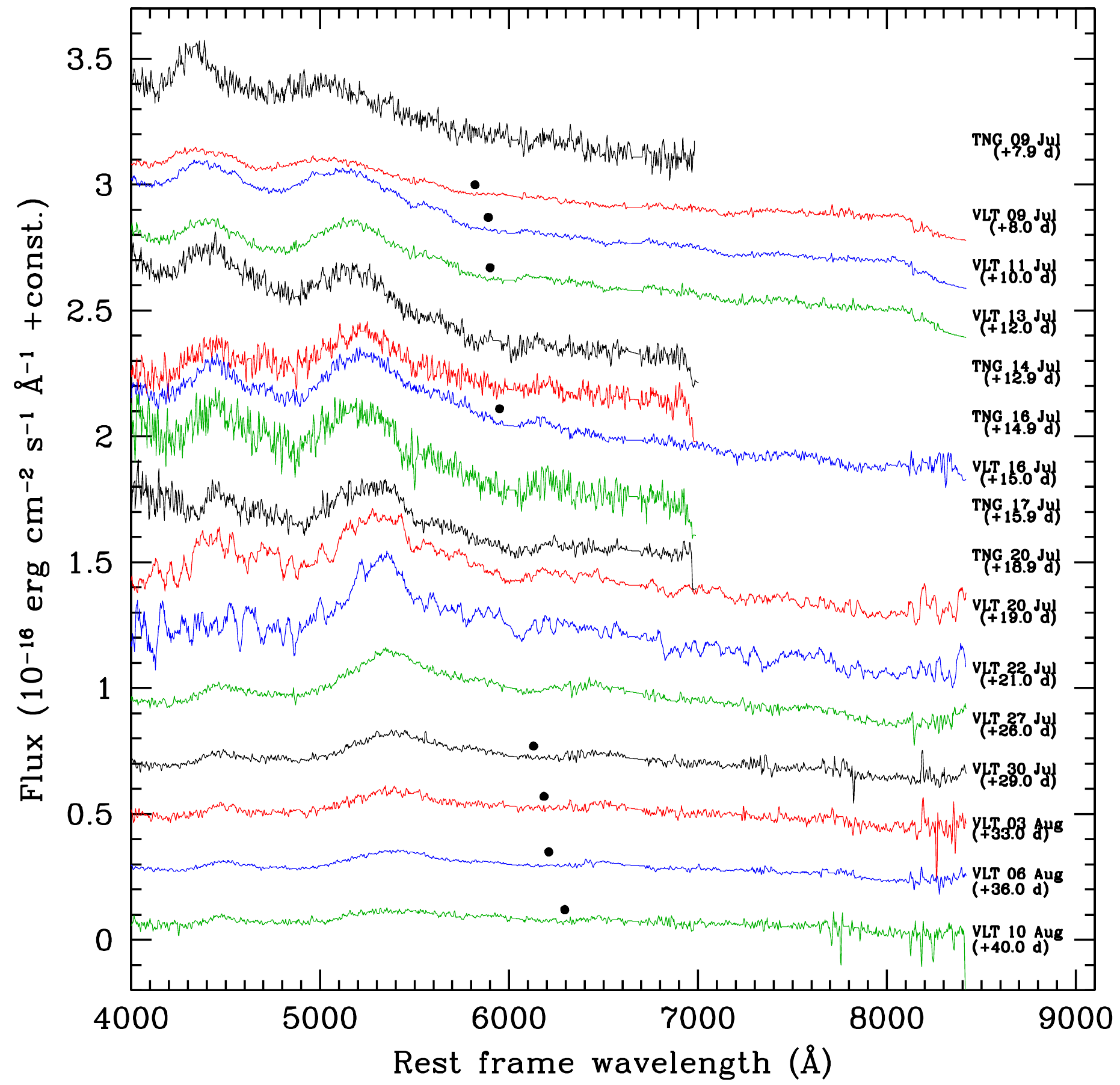}
\caption{Sixteen dereddened, host- and afterglow-subtracted
  TNG and VLT spectra.  For clarity, the spectra have been vertically
  shifted by $2\times 10^{-17}$ erg cm$^{-2}$ s$^{-1}$ \AA$^{-1}$,
  with the earliest one (TNG, Jul. 9) at top and the latest one, not
  shifted, at the bottom. All VLT spectra are smoothed with a boxcar of
  $8$ \AA, with the exception of those from Jul 20 and 22  (that have the
  lowest S/N ratio), which are smoothed by $25$\AA. All TNG spectra
  are smoothed with a boxcar of $7$ \AA.  The time intervals are
  computed from the GRB explosion time. Black filled circles mark
    the position of the SiII6355 feature, used to determine the
    photospheric velocity of the ejecta (see Sect. 4.4).}
\end{figure*}

\subsection{Optical spectra}

First, all the VLT and TNG optical spectra of SN 2013dx were
dereddened using the Galactic value $E(B-V)=0.04$ mag toward its line
of sight (Schlegel, Finkbeiner, \& Davis 1998).  The intrinsic
extinction appears to be negligible both from X-ray data and from
analysis of the X-ray-to-optical spectral energy distribution.  Then,
we estimated the host galaxy contribution. Since SN 2013dx was still
bright at the time of our last observation (10 Aug.) and later on its
field became not observable anymore because of solar constraints, we could
not acquire a reliable spectrum to subtract the host galaxy
contribution from our data.  Therefore, we considered the SDSS
magnitudes of the host galaxy, $u=24.42\pm 0.96$, $g= 23.81\pm0.85$,
$r = 23.01\pm0.24$ $i=23.22\pm0.39,$ and $z=23.04\pm0.59$, and noted
that after they were reduced to rest-frame, they were consistent with the
normalized template of a star-forming galaxy with moderate intrinsic
absorption ($E(B-V) <0.10$, Kinney et al. 1996).  Then we subtracted
this rescaled template from our dereddened fluxes.  The host galaxy
contributes about 15\% of the total flux, and possibly less, because
not all its light is captured in the slit when acquiring the SN
spectra.

Finally, we subtracted the afterglow contribution. The afterglow light
curve can be modeled with a broken power law, as shown in Singer et
al. (2013). We thus adopted their first decay index ($\alpha_1=0.57$)
and their temporal break $t_{b}=1.17$ days. However, the second decay
index $\alpha_2$ does not take into account the emerging SN
contribution. $\alpha_2=1.85$ is the lowest index for which the
afterglow is not oversubtracted in our early time photometry, while
$\alpha_2=2.5$ is the highest index allowed by the closure relations
linking spectral and temporal indices (Zhang \& M\'esz\'aros
2004). Thus, we chose an average $\alpha_2 = 2.2$. The adopted
spectral index is $\beta_{\nu}=0.7$ (Singer et al. 2013). Even
assuming the two extreme decay indices, the difference in the
afterglow contribution can only be appreciated in the first three spectra, and it is lower than 20\% in the worst case.

Figure 3 shows our sixteen dereddened, host- and afterglow-subtracted
TNG and VLT spectra. The spectra are shown starting from $4000$\AA\ in
the rest-frame, because the flux calibration becomes unreliable at
shorter wavelengths. 

In Fig. 4 we compare in the rest-frame 11 of our spectra with those of
SN 1998bw at comparable phases after explosion (Patat et
al. 2001). Each SN 1998bw spectrum is scaled in flux by an arbitrary
constant to find the best match with our spectral dataset.  In most
pairs of spectra the same continuum shape and broad absorption
features can be seen, even if some diversities are present (at
  t$<20$ d the $4400$\,\AA\, pseudo-emission peak is completely absent in
  1998bw; the pseudo-peak around $6300$\,\AA\, in 1998bw is absent in
  2013dx; the strong absorption around $7000$\,\AA\ in 1998bw is much
  less pronounced in 2013dx at t$<15$ d). This fact leads us
to classify
SN 2013dx as a broad-line type Ic SN, that is, one with a highly stripped
progenitor (no hydrogen or helium left before explosion), similar
to those of previously studied GRB- and XRF-SNe (Mazzali et
al. 2006a,b).  Other GRB-associated SNe, such as 2003dh, 2006aj, and
2010bh, do not compare spectrally as convincingly with SN 2013dx.


\begin{figure}
\centering
\includegraphics[angle=0,width=9cm]{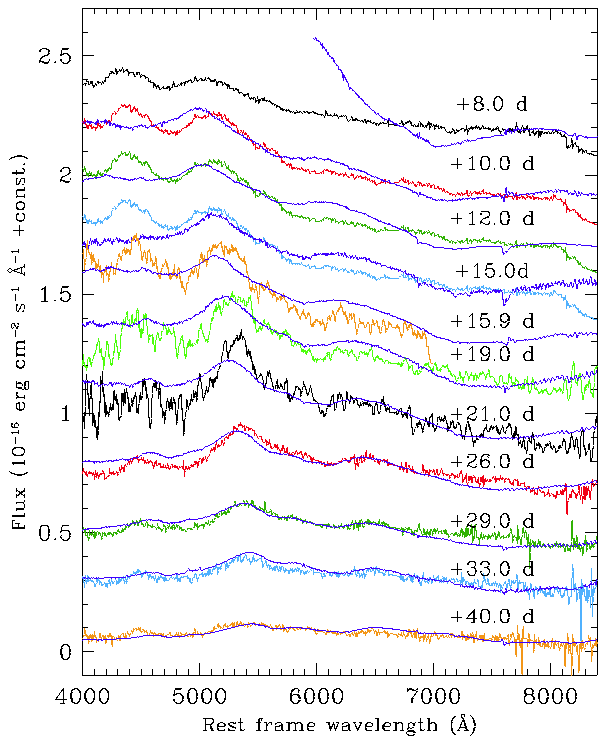}
\caption{
 Eleven  spectra of SN 2013dx (various colors) and 1998bw (blue only) at comparable 
 rest-frame phases, marked in days with respect to the explosion time of GRB130702A. 
As in Fig. 3,  the SN 2013dx spectra have been
  smoothed and vertically shifted by $2\times 10^{-17}$ erg cm$^{-2}$
  s$^{-1}$ \AA$^{-1}$, with the earliest one at the top and the latest
  one, not shifted, at the bottom.}
\end{figure}

Of the broad-lined SNe that are not accompanied by GRBs/XRFs, the type Ic SN
2010ah shows a remarkable spectral similarity.  In Fig. 5 we show the
comparison of the models that best describe the two available spectra
of SN2010ah (Mazzali et al. 2013) with the spectra of SN2013dx
(corrected as detailed above) taken at comparable phases.  

The agreement between the two SNe is quite good at $4500 \ls
\lambda \ls 6000$ \AA, where the most relevant absorption features
are located.  At wavelengths shorter than $4500$ \AA\, some residual
contamination from the afterglow or a shock breakout component
(e.g., Campana et al. 2006; Ferrero et al. 2006) might still be present
in the SN2013dx spectrum, while at wavelengths higher than $6000$
\AA\, the model shown for SN2010ah clearly does not match SN 2013dx accurately.  We note that both SNe show a SiII absorption
line ($\sim6000$ \AA) and O I absorption line ($\sim7300$ \AA), but
they are weaker in SN2013dx than in SN2010ah. This may suggest a
reduced abundance of silicon and oxygen in the former SN.



\begin{figure}
\centering
\includegraphics[angle=0,width=9cm]{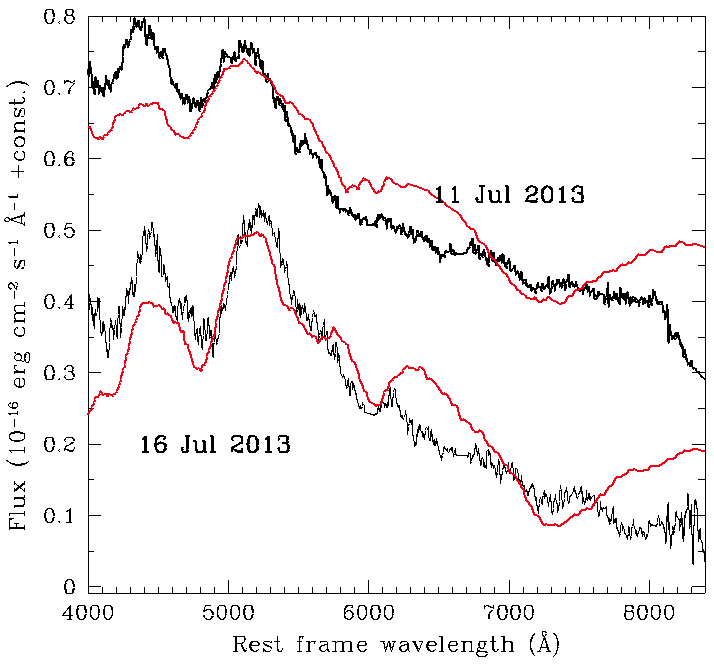}
\caption{
  Spectra of SN2013dx (black) taken on 11 and 16 July 2013, in
  rest-frame.  For comparison, the models of the two available spectra
  of SN2010ah, taken at comparable phases after SN explosion, are
  shown in red.  As in Fig. 3, the Jul 11 spectrum has been vertically
  shifted by $3\times 10^{-17}$ erg cm$^{-2}$ s$^{-1}$ \AA$^{-1}$ with
  respect that of Jul 16.  }
\end{figure}

\subsection{Bolometric light curve}

We computed a bolometric light curve in the range $3000--10000$ \AA\
from our multicolor light curves.  After correcting the photometric
data in an analogous way as done for the spectra (Galactic extinction,
host galaxy, and afterglow contribution, see Sect. 4.2) and after
applying k-corrections using our VLT spectra (that cover a wider
wavelength range than the TNG spectra), we splined the residual
monochromatic light curves, which should represent the supernova
component, and the broad-band flux at each photometric observation
epoch was integrated. The flux was linearly extrapolated blueward of
the U-band flux down to 3000 \AA\ and redward of the I-band flux to
$10000$ \AA. The result is reported in Fig. 6. The errors associated
with the photometry, host galaxy template, and afterglow were
propagated and summed in quadrature. The figure also shows the
bolometric light curves of SN 1998bw (Patat et al. 2001), SN 2006aj
(Pian et al. 2006), SN 2010bh (Bufano et al. 2012), SN 2003dh (Hjorth
et al. 2003), SN 2003lw (Mazzali et al. 2006a), SN 1994I (Richmond et
al. 1996) and SN 2012bz (Melandri et al. 2012).


\begin{figure}
\centering
\includegraphics[angle=0,width=9.2cm]{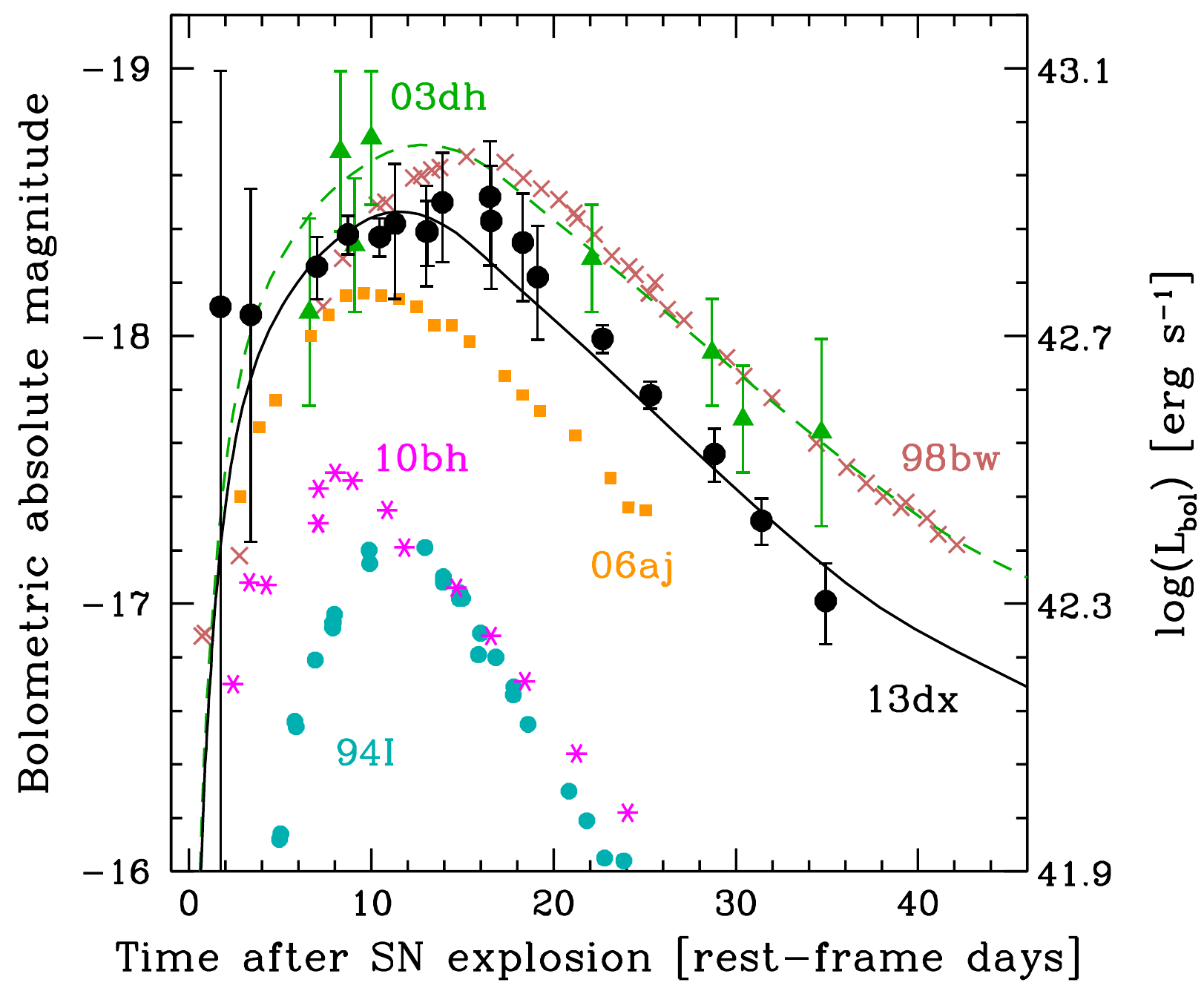}
\caption{Bolometric light curve of SN 2013dx (large circles) compared 
with those of other core-collapse type Ic SNe associated with a GRB or XRF 
(SNe 1998bw, crosses; 2003dh, triangles; 2006aj, squares; 2010bh, 
asterisks) and not associated with any detected high-energy event (SN1994I, 
small circles).   Before day 20, several points of SN2013dx are affected by 
larger uncertainties than at later epochs, despite the higher brightness,  
because they are derived from the TNG photometry that is somewhat more
noisy than the VLT photometry.  The dashed line represents the
model light curve for SN 2003dh, corresponding to the synthesis of
0.35 M$_\odot$ of $^{56}$Ni, while the solid curve, which best
reproduces the bolometric flux of SN 2013dx, was obtained by scaling
down the above model curve by 25\% and compressing it by 10\%, which
corresponds to a synthesized $^{56}$Ni mass of $\sim$0.2 M$_\odot$.
}
\end{figure}

\subsection{Photospheric velocities}

To estimate the velocity of the SN ejected material, we started
out from the general similarity between our spectra and those of SN1998bw. 
Following Patat et al. (2001, see their Fig. 5), we identified
the {\ion{Si}{II}$\lambda$6355} feature and followed its redward shift
along our spectra. In the spectra acquired between 17 and 
22 July, the cosmological redshift and the blueshift due to the
photospheric expansion combine to cause this absorption doublet to overlap
with the telluric feature at 6870 \AA. For these spectra, the position of
the {\ion{Si}{II}$\lambda$6355} cannot be measured.


\begin{figure}
\centering
\includegraphics[angle=0,width=9cm]{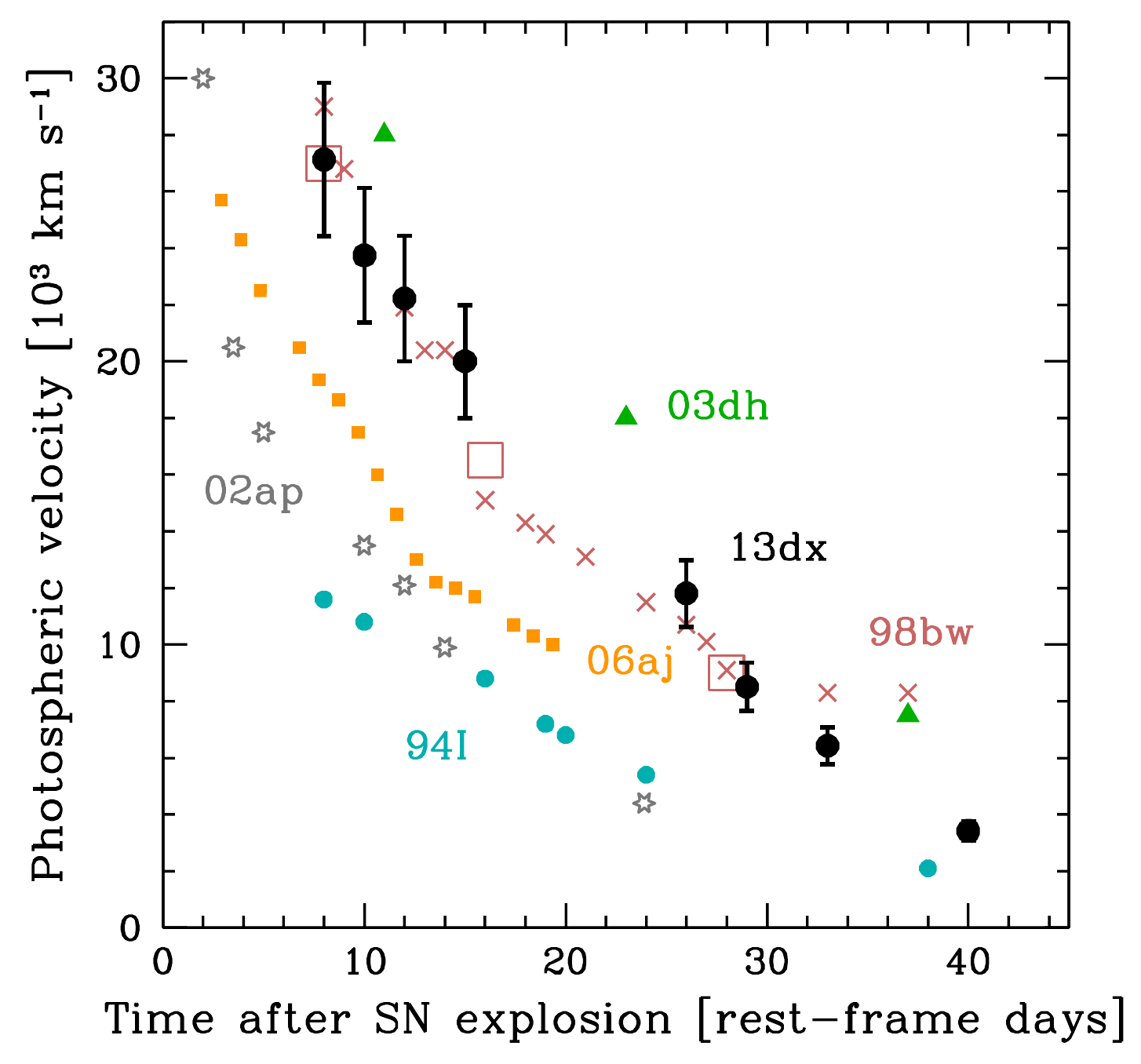}
\caption{Temporal evolution of the photosphere expansion 
velocity of SN 2013dx (large circles) measured directly on the
spectra from the position of the minimum of the {\ion{Si}{II}$\lambda$6355} 
absorption line.  For comparison we report the photospheric velocities of 
other core-collapse Ic SNe  (1994I, small circles; 2002ap, stars; 2003dh, triangles; 
2006aj, small filled squares), as derived from spectral models (Sauer et al. 2006; Mazzali et al. 
2002; 2003; 2006b). For SN1998bw we report 
both the velocities derived from direct measurements (crosses, Patat et al. 2001) 
and those derived from models  (large open squares, Nakamura et al. 2001).}
\end{figure}

The time evolution of the photosphere expansion velocity as deduced
from the minimum of the absorption trough of the
\ion{Si}{II}$\lambda$6355 line is shown in Fig. 7 (large 
circles). The velocity decline extends from $\sim 2.7\times 10^{4}$ km
s$^{-1}$ at $8$ rest frame days from the explosion to $\sim 3.5\times
10^{3}$ km s$^{-1}$ at $40$ days. Figure 7 also displays the
photospheric velocities of other SNe, in particular that of SN 1998bw,
computed by Patat et al. (2001) using the same procedure as that
adopted by
us. The photospheric velocities of SN 2013dx and SN 1998bw agree very
well, although at very late time the latter retains a
slightly higher velocity expansion (see also Fig. 4 of Patat et
al. 2001). Comparing the expansion velocity of SN 2013dx with that of
other SNe, we find that the former is higher than that of SN 1994D
(Patat et al. 1996), SN 1994I (Millard et al. 1999), SN 1997ef (Patat
et al. 2001) and SN 2006aj (Mazzali et al. 2006), but not as extreme
as that of SN 2003dh (Hjorth et al. 2003).

\subsection{Physical parameters of SN 2013dx}

The bolometric light curve of SN 2013dx shows that the peak
luminosity of this SN is intermediate between those of the nearby GRB-SNe
1998bw and 2003dh on one hand and those of the two XRF-SNe 2006aj and
2010bh (Fig. 6) on the other. The spectrum of SN 2013dx is much more
similar to that of the most energetic Ic SNe, and especially to SN
2010ah, which, among the broad-lined Ic SNe not accompanied by a
GRB/XRF, is the most spectroscopically similar to SN 1998bw (Mazzali
et al. 2013), although because of its lower ejecta mass, it has a
significantly lower kinetic energy ($1.2 \times 10^{52}$ erg vs $5
\times 10^{52}$ erg inferred by Nakamura et al. 2001 for SN 1998bw).

The similarity of the light-curve shape of SN 2013dx to the shapes
of those of SN
1998bw and SN 2003dh and the spectral resemblance to SN 1998bw and SN
2010ah led us to adopt these three previously known and well-studied
SNe as "templates" for estimating the physical parameters of SN
2013dx through the relationships that link the SN ejecta mass $M_{ej}$
and kinetic energy $E_K$ to the observables, that is, to the width of the
bolometric light curve and the photospheric velocities (Arnett 1982,
see Sect. 4.1 in Mazzali et al. 2013).  This method, described in
detail by Mazzali et al. (2013; see also Valenti et al. 2008; Walker
et al. 2014), should be applied with caution when only few SNe are
available that provide a good match and the data coverage of both
target and templates is not excellent.  However, when the
observational information is adequate, the outcome satisfactorily agrees with the results of modeling based on radiative
transport: for SN 2010ah, which has a well-sampled light curve, but only
two spectra taken around maximum light, the physical parameters obtained
with the two approaches differ by no more than 25\% (Mazzali et
al. 2013).

The dataset presented here for SN 2013dx is detailed and rich, and we
can compare it with as many as three template SNe with good data for
which models were developed (Nakamura et al. 2001; Deng et al. 2005;
Corsi et al. 2011; Mazzali et al. 2003; 2006a; 2013). This allows us
to use the rescaling method to describe the physics of SN 2013dx with
good accuracy. Developing a dedicated model is beyond the
scope of this paper, but will be presented in the future.

Based on the similarity of the light-curve shape of SN2013dx and
SN2003dh, we adapted the model curve of 2003dh (Mazzali et al. 2003;
Deng et al. 2005) by compressing it by 10\% in time and dimming it by
25\% in flux.  We thus obtained the synthetic curve reported in Fig. 6
(solid line).  Since SN 2003dh produced an estimated $^{56}Ni$ mass of
$\sim$ 0.35 M$_\odot$ (Mazzali et al. 2006a), the 25\% flux dimming
corresponds to a $^{56}Ni$ mass of 0.26 M$_\odot$.  However, this is
further reduced by the effect of the 10\% faster temporal evolution of
SN 2013dx with respect to SN 2013dx.  By taking this into account and
assuming $^{56}Ni$ decay dominates at early times (i.e., neglecting the
contribution of $^{56}Co$ decay), we obtain a further reduction of
20\%, meaning that the final estimated $^{56}Ni$ mass synthesized by SN 2013dx
should be $\sim$0.2 M$_\odot$.

From a spectroscopic point of view, SN2013dx is a close analog of
the GRB-SN prototype SN1998bw and a very close analog of the
broad-lined type Ic SN2010ah, although in the latter case the
comparison is limited to only two spectra.  This suggests that the
photospheric velocities are similar and that the kinetic energy of
SN2013dx may then just simply scale like the ejecta mass.

Using the observed light curve width $\tau$ and measured photospheric
velocities of SNe 1998bw, 2003dh, 2013dx (Figs. 6 and 7) and SN 2010ah
(Mazzali et al. 2013), we applied the scaling relationships to
SN2013dx and to each of the adopted template SN.  Then we averaged
the results and found for SN2013dx $M_{ej} = 7 \pm 2$ M$_\odot$ and
$E_K = (35 \pm 10) \times 10^{51}$ erg.  Our errors on these physical
quantities reflect conservatively the empirical nature of the method
and its uncertainties (for instance, the fact that the phases at which
the measured photospheric velocities are compared are never identical
for the target and the template SN). The progenitor of SN 2013dx is
thus probably $25-30 M_{\odot}$, which is $15- 20\%$ less massive than
that estimated for SN 1998bw ($30-35 M_{\odot}$, Maeda et al. 2006)

\subsection{Field of SN2013dx}

The field of SN 2013dx shows a bright galaxy southward of the transient,
at $\sim 8"$ from the SN position (see Fig. 8). This could have been
the SN host, since the faint source ($R=23.01$) at $0.6"$ from SN
2013dx was initially misclassified as a star in the SDSS (Singer et
al. 2013). For this reason, Leloudas et al. (2013) obtained a spectrum
of this galaxy, reporting a $z=0.145$, which is the same as that
of the
transient (Mulchaey et al. 2013a,b; D'Avanzo et al. 2013d). Kelly et
al. (2013) performed an extensive study of this galaxy and its outskirt,
concluding that the SN 2013dx host galaxy is a dwarf satellite of the
bright one.

Many more galaxies are present in the SN 2013dx field. Kelly et
al. 2013 noted that many of them have an SDSS photometric redshift
compatible with that of the host and the bright galaxy. In addition,
they also pointed out that the SDSS spctroscopic galaxy survey
targeted five galaxies brighter than $17.7$ within 15' from SN 2013dx,
their redshift being close to $z=0.145$.

We decided to use different slit position angles during our VLT
spectroscopic campaign. This allowed us to study both the
spectroscopic evolution of the SN and to determine the redshift of
more nearby galaxies. The different position angles adopted are
described in Table 5, and the corresponding slit orientations are
displayed in Fig. 8. This figure also marks with numbers the $14$
galaxies that we placed in the slit in our observations ("S" marks a star).

We determine the redshift of all these galaxies, although for three of
them the determination is not completely certain because it relies on
faint absorption or emission features. Table 6 illustrates the
redshift of the galaxies, which are numbered as in Fig. 8. The
uncertain redshifts are marked with "?", while the last two columns
of the table list the emission and absorption features on which the
redshift determination is based. The spectra of our $14$ field
galaxies are shown in appendix A1.

It is interesting to note that $9$ out of $14$ galaxies lie within
$0.03$ from $z=0.145$, that is, the redshift of  SN 2013dx, and one
more is within $0.2$. These $\text{ten}$ galaxies are marked in red in Fig. 8,
while the remaining $\text{\\ four}$ are marked in blue.  In particular, we confirm
the photometric redshifts reported for two of the field galaxies in
Fig. 2 of Kelly et al (2013), their sources S3 and S5. This is clear evidence that SN 2013dx occurred in a group or a small cluster
of galaxies. This conclusion is strengthened by the fact that many
more galaxies in the SDSS, falling just outside the field of view of
our images, have a spectroscopic or photometric redshift consistent
with $z=0.145$.  In detail, since the angular separation among the two
farthest galaxies with the same redshift in the field is $\sim 3'$,
the physical extent of the group at $z=0.145$ is $\sim 600$ kpc. This
is the first SN associated with a long GRB detected in such an
environment (and, to our knowledge, the first long GRB in
general). For the SN 1998bw environment, a candidate cluster was first
claimed (Duus \& Newell 1977), but then not confirmed by spectroscopic
analysis (Foley et al. 2006). For SN 2012bz, associated with
GRB\,120422A, two objects have been reported at the redshift of the
GRB (Schulze et al 2014). However, this looks like an interacting
system consisting of two or possibly three galaxies, and not like a
group.

Intuitively, one may argue that the probability of hosting a GRB
should be higher in the largest galaxy of the cluster, where the
maximum star-formation occurs. However, the dwarf galaxies have
comparatively high or higher specific star-formation rates, which
might bias the probability of hosting a GRB in
their favor.
Indeed, studies on complete samples of GRBs (see, e.g., Perley et
al. 2014; Vergani et al. 2014) show that at $z<1$ long GRBs strongly
prefer low-mass galaxies (confirming previous studies on incomplete
samples). This is probably because GRB progenitors are
more easily developed in low-metallicity environments, which possess a
high specific SFR.


\begin{figure}
\centering
\includegraphics[angle=0,width=9cm]{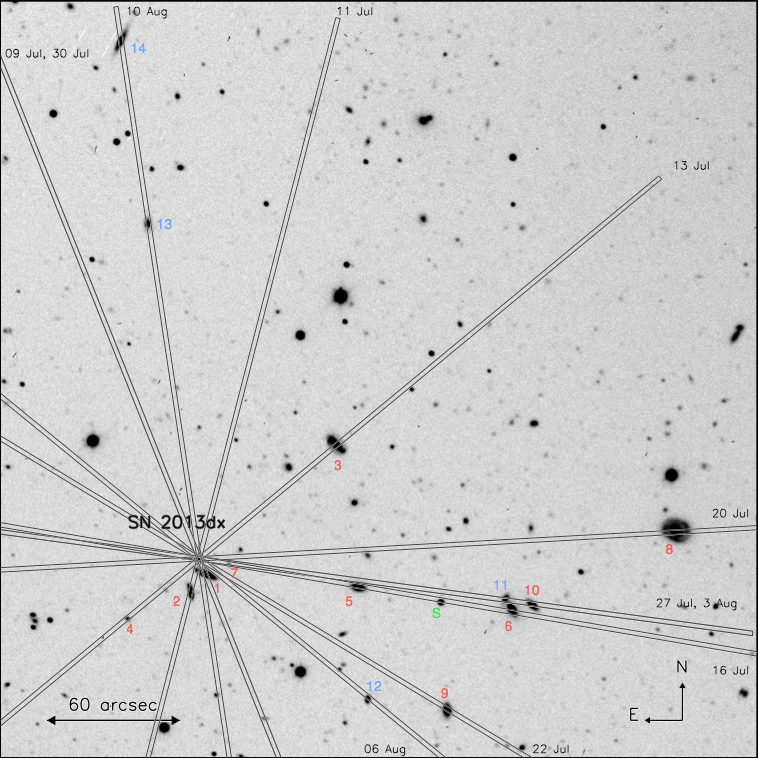}
\caption{Field of SN 2013dx with the slit positions used for
 FORS2   spectroscopy overimposed. Each slit position is
  identified by its observation date. Numbers mark the galaxies for
  which spectroscopic detection was secured. Red numbers refer to
  galaxies with redshift close to that of the SN 2013dx host galaxy
  ($z=0.145)$. Blue numbers refer to galaxies not related to the
  SN.  The green "S" marks a field star included in the slit  on 16  July.}
\end{figure}

\begin{table*}[ht]
\caption{\bf Observation dates, redshifts, and detected lines for the galaxies in the field of SN 2013dx}
\centering
{\footnotesize
\smallskip
\begin{tabular}{|l|c|c|c|c|c|}
\hline

Galaxy & obs date     & exp. time (s) & redshift   & emission lines                                            & absorption lines        \\
\hline
\hline
1      & 09 Jul   & $1800$            & $0.145$  & H$\alpha$, SII, H$\beta$, OIII, OII, OI                       & H\&K Ca, Gband MgI, NaI \\
\hline
2      & 11 Jul   & $900$             & $0.145$  & H$\alpha$, SII, H$\beta$, OIII, OII, OI, ArIII, NeIII         & H\&K Ca, Gband MgI, NaI \\
\hline
3      & 13 Jul   & $900$             & $0.147$  & H$\alpha$, SII, H$\beta$, OIII, OII                           & H\&K Ca, Gband MgI \\
\hline
4      & 13 Jul   & $900$             & $0.142?$  & none                                                      & H\&K Ca, Gband MgI, NaI, H$\alpha$, H$\beta$, OIII \\
\hline
5      & 16 Jul   & $900$             & $0.143$  & none                                                      & H\&K Ca, Gband MgI, NaI, H$\alpha$, H$\beta$ \\
\hline
6      & 16 Jul   & $900$             & $0.146$  & OII (w)                                                   & H\&K Ca, Gband MgI, NaI, H$\alpha$, H$\beta$ \\
\hline
7      & 16 Jul   & $900$             & $0.148?$ & H$\alpha$ (w), H$\beta$ (w), OII (w)                          & MgI \\
\hline
8      & 20 Jul   & $900$             & $0.163$  & OII                                                       & H\&K Ca, Gband MgI, NaI, H$\beta$, ArIII \\
\hline
9      & 22 Jul   & $900$             & $0.144$   & OII (w)                                                   & H\&K Ca, Gband MgI, NaI, H$\alpha$, H$\beta$ \\
\hline
10     & 27 Jul   & $900$             & $0.142$  & none                                                      & H\&K Ca, Gband MgI, NaI, H$\alpha$, H$\beta$ \\
\hline
11     & 27 Jul   & $900$             & $0.336?$   & H$\alpha$(w), H$\beta$ (w), OII, HeI (w), HeII (w)            & H\&K Ca, Gband \\
\hline
12     & 06 Aug   & $3600$            & $0.336$   & H$\alpha$, H$\beta$, OII, OIII, HeII, MgII(w), SII(w), OI (w) & H\&K Ca, Gband \\
\hline
13     & 10 Aug   & $1800$            & $0.423$  & none                                                      & H\&K Ca, Gband, MgI \\
\hline
14     & 10 Aug   & $1800$            & $0.237$  & H$\alpha$, OII, OIII                                        & H\&K Ca, Gband, MgI, NaI, HeI \\
\hline
\hline

\end{tabular}
}
\end{table*}

\section{Conclusions}
 
We have presented an extensive and sensitive ground-based
observational campaign on the SN associated with GRB\,130702A at $z
=0.145$, one of the nearest GRB-SNe detected so far.  Its relative
proximity guaranteed the construction of a nice dataset with a good
S/N. The properties of SN 2013dx are similar to those of previous GRB-
and XRF-SNe: the peak luminosity is intermediate between those
of GRB-SNe and XRF-SNe, and the photospheric velocities are more
similar to those of GRB-SNe.  Accordingly, the physical parameters of
SN 2013dx, derived with the empirical method based on the rescaling of
the quantities known for other SNe, are similar to those determined
for the previous GRB-SNe, but somewhat lower than those of SN 1998bw:
we estimate a synthesized $^{56}Ni$ mass of $\sim$ 0.2 M$_\odot$, an
ejecta mass of $M_{ej} \sim 7$ M$_\odot$, and a kinetic energy of $E_K
\sim 35 \times 10^{51}$ erg.

Furthermore, we performed a study of the SN 2013dx environment through
spectroscopy of the field galaxies close to the host of GRB\,130702A.
We find that $65\%$ of the observed targets have the same redshift as
SN 2013dx, indicating that this is a group of galaxies. This
represents the first report of a GRB-SN association taking place in a
galaxy group or cluster.

\begin{acknowledgements}

  We thank an anonymous referee for several helpful comments. We are
  grateful to the ESO Director for awarding Discretionary Time to this
  project.  We thank S. Valenti for several helpful discussions and
  F. Patat for providing the photospheric velocities for SN 1998bw.
  We thank the TNG staff, in particular G. Andreuzzi, L. Di Fabrizio,
  and M. Pedani, for their valuable support with TNG observations, and
  the Paranal Science Operations Team, in particular H. Boffin,
  S. Brillant, C. Cid, O. Gonzales, V. D. Ivanov, D. Jones,
  J. Pritchard, M. Rodrigues, L. Schmidtobreick, F. J. Selman,
  J. Smoker and S. Vega. The Dark Cosmology Centre is funded by the
  Danish National Research Foundation. VDE acknowledges partial
  support from PRIN MIUR 2009. This research was partially supported
  by INAF PRIN 2011, PRIN MIUR 2010/2011, and ASI-INAF grants
  I/088/06/0 and I/004/11/1. FB acknowledges support from FONDECYT
  through Postdoctoral grant 3120227 and from Project IC120009
  "Millennium Institute of Astrophysics (MAS)" of the Iniciativa
  Cientifica Milenio del Ministerio de Economia, Fomento y Turismo de
  Chile. D.M. acknowledges the Instrument Center for Danish
  Astrophysics for support. The spectra are publicly available on
  WISeREP - http://wiserep.weizmann.ac.il.

 \end{acknowledgements}

\appendix
\section{SN 2013dx field galaxy spectra}

In this appendix we report the $14$ spectra of the galaxies in the
field of SN 2013dx (see also Sect. 4.6).

\begin{figure*}
\centering
\includegraphics[angle=0,width=19cm]{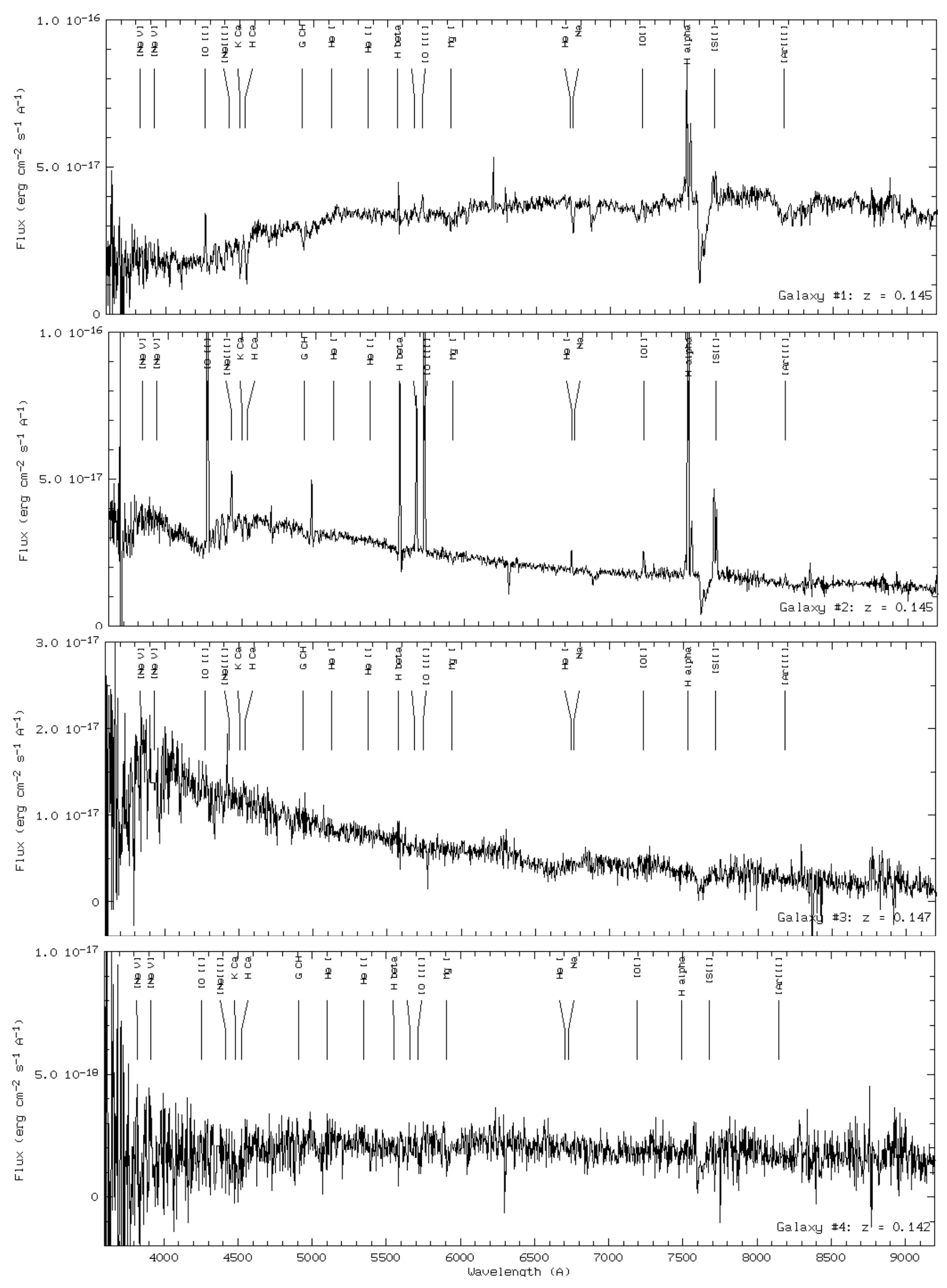}
\caption{Spectra of the field galaxies of SN 2013dx. Wavelengths
  are in observer frame and galaxies are numbered as in Table 6. and
  Fig. 8.}
\end{figure*}

\begin{figure*}
\centering
\includegraphics[angle=0,width=19cm]{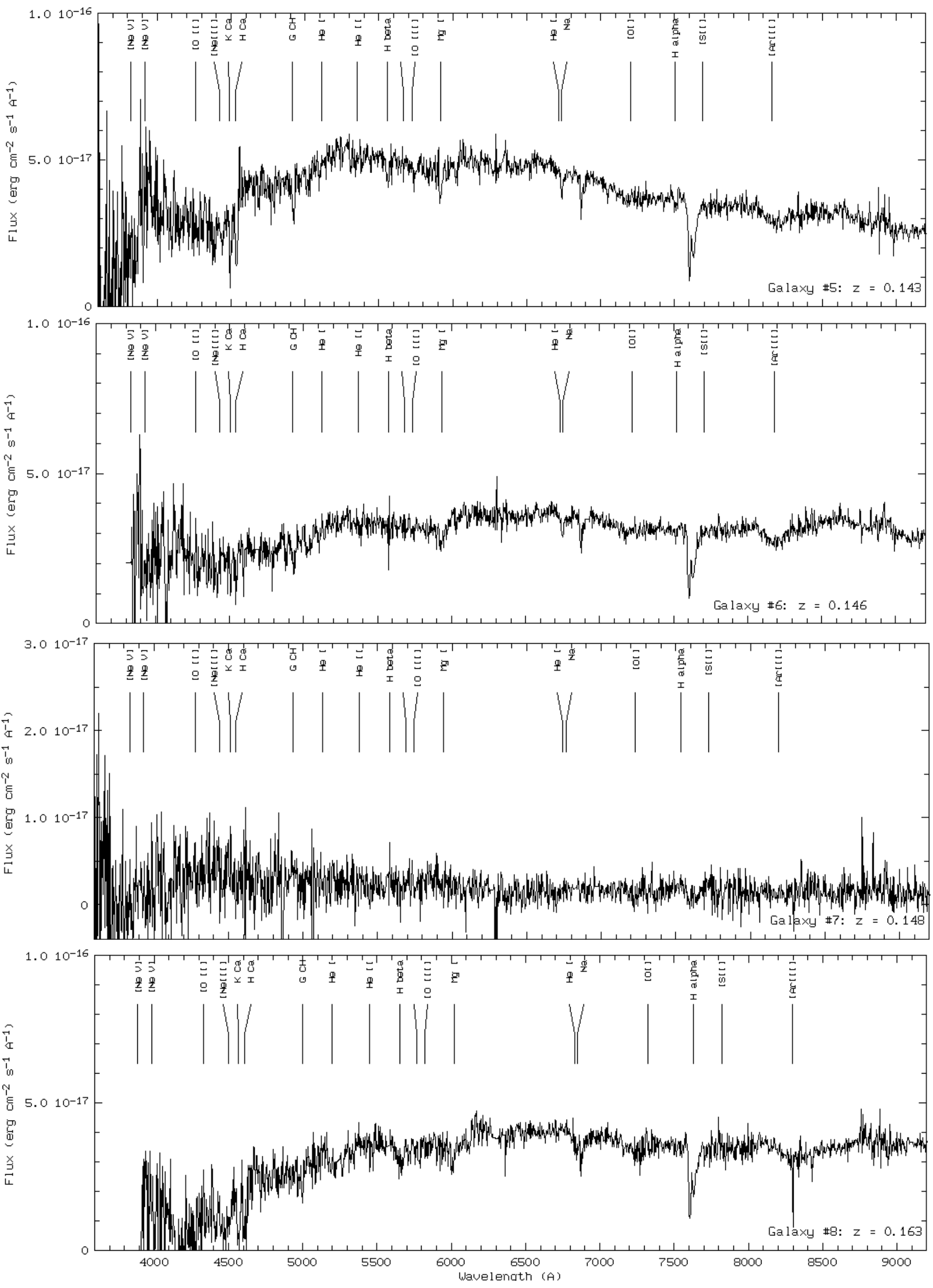}
\caption{- continued}
\end{figure*}

\begin{figure*}
\centering
\includegraphics[angle=0,width=19cm]{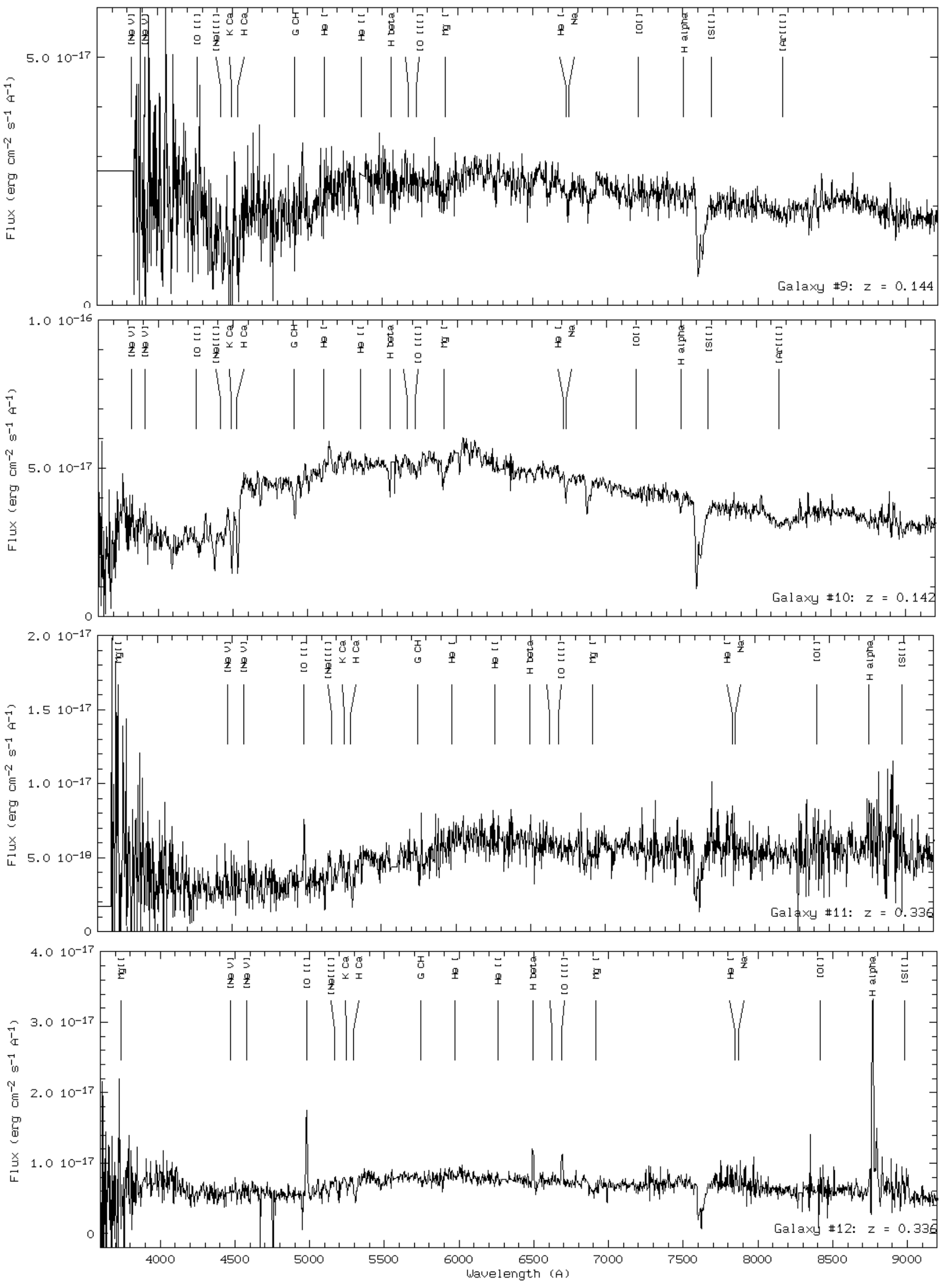}
\caption{- continued}
\end{figure*}

\begin{figure*}
\centering
\includegraphics[angle=0,width=19cm]{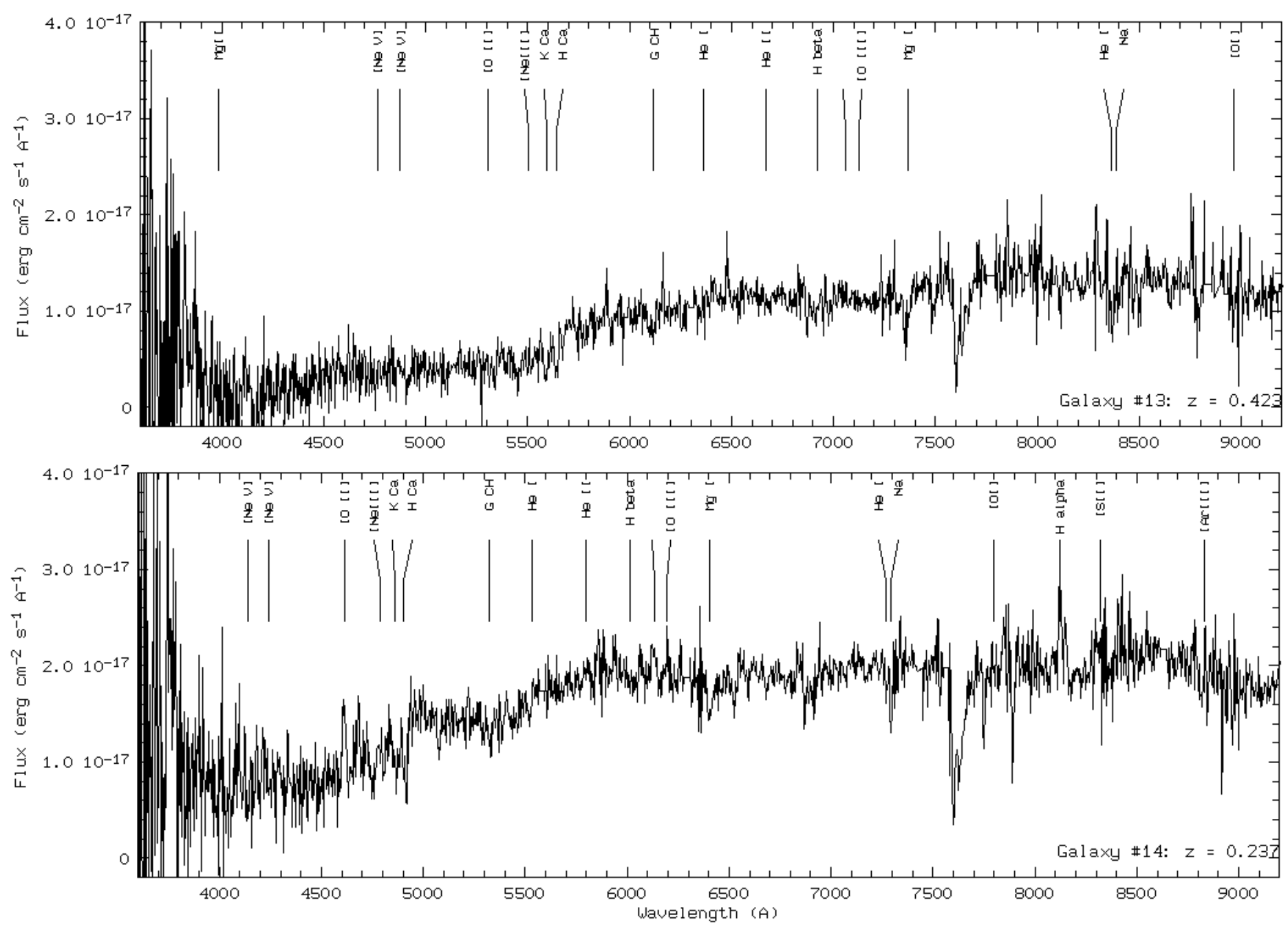}
\caption{- continued}
\end{figure*}

\end{document}